\documentclass{trans2-l}

\newtheorem{theorem}{Theorem}[section]

\theoremstyle{definition}

\theoremstyle{}
\newtheorem{remark}[theorem]{Remark}

\numberwithin{equation}{section}

\begin{document}

\title{Feynman integral for functional  Schr\"{o}dinger equations }

\author{Alexander Dynin}
\address{Department of Mathematics, Ohio State University,
Columbus, Ohio 43210, USA}
\email{dynin@math.ohio-state.edu}

\subjclass{Primary 58D30  ; Secondary 7G30, 81Q05}
\date{}

\dedicatory{This paper is dedicated to M. I. Vishik and
his  Seminar at the Moscow State University.}

\keywords{Constructive quantum field theory, Infinite-dimensional
pseudodifferential calculus, Functional Schr\"{o}dinger equations,
Feynman Integrals. }

\begin{abstract}
We consider functional Schr\"{o}dinger equations associated with a wide class
   of    Hamiltonians  in all Fock
    representations of the bosonic canonical commutation relations, in
    particular the Cook-Fock,   Friedrichs-Fock, and  Bargmann-Fock models.
An infinite-dimensional symbolic calculus allows to prove
the convergence of the corresponding Hamiltonian Feynman  integrals
for propagators of coherent states.
\end{abstract}

\maketitle

\specialsection*{Dedication}
The roots of my appreciation of M. I. Vishik and  his PDE
     Seminar go back to the golden 1960's
of Moscow mathematics (cf. \cite{Zdr93}).

In the Spring  of 1960  A. Volpert gave a lecture
     at the Moscow Gelfand seminar about his index formula of
     elliptic boundary problems  for PDE systems in two variables.
     His approach was based on the Muskhelishvili index formula of
     one-dimensional singular integral operators.

I. M. Gelfand interpreted the Volpert index formula in terms of characteristic
cohomological classes and suggested to generalize it to elliptic boundary
     problems in higher dimensions  and  on compact  manifolds using
a homotopy of the elliptic data.

The   Vishik  seminar began  in the Spring of  1961   with a response to the
      Gelfand challenge. It started with a review of  still recent
      Calderon-Zygmund algebra of
multivariable singular integral operators and their applications to
the Cauchy problem for elliptic PDE's. Soon the Calderon-Zygmund algebra
was extended to the algebra of singular
integro-differential operators (see \cite{Dyn61}).
The elliptic homotopy of the extended algebra is much simpler
because its symbols are continuous rather than polynomial in cotangent
directions. Indeed,  this extension was essential for
     the first general solution of the Gelfand index
     problem,  given by M.~Atiyah and I.~Singer \cite{AS63}.

As stated in  \cite{KN65}, the  Kohn-Nirenberg
pseudodifferential operator algebra  is a refinement of the algebra of singular
integro-differential operators.

Later M.~I.~Vishik  initiated  the infinite-dimensional
pseudodifferential operator theory (cf. \cite{Vis71}).

This article is a contribution in this direction.

\section{Introduction}

\subsection{Analytical aspects of quantum field theory}
We begin with E.Witten's remarks (\cite{Wit00}, p.346)
\begin{quotation}
Mathematically, quantum field theory involves integration, and
elliptic operators, on infinite-dimensional spaces. Naive attempts
to formulate such notions in infinite dimensions lead to all sorts of
trouble. To get somewhere, one needs the very delicate constructions
considered in physics, constructions that at first sight look rather
specialized to many mathematicians. For this reason, together with
inherent analytical difficulties that the subject presents,
rigorous understanding has tended to lag behind development of physics.
\end{quotation}

There are three basic analytical formulations of  quantum field theory
{\cite{BS80}. They describe either  a classical evolution of quantum
fields, or  a quantum evolution of classical fields.

We sketch the case of neutral bosons of positive mass $\mu$ where
the classical fields  are represented by
      real  functions $\phi_{t}(x)$ on the Euclidean space
     $\mathbf{R}^{d}$.

     The classical Lagrangean is a real  function
     $L(\phi_{t}, \nabla\phi_{t}, \dot{\phi}_{t})$ of $\phi_{t}$ and its
     space and time derivatives. E.g., for  self-interacting bosons
     \begin{equation}
      L=\frac{1}{2}[\dot{\phi}_{t}^{2}-(\nabla\phi_{t})^{2}-\mu^{2}\phi_{t}^{2}]
      -V(\phi_{t}),
\end{equation}
where $V$ is a scalar  potential. (If $V=0$ then the boson field is
     free.)

     The classical Hamiltonian is a real function
     $H(\phi_{t}, \nabla\phi_{t},\pi_{t})$, where
     $\pi_{t}=\partial/\partial\dot{\phi}_{t}$.

     In our example
     \begin{equation}
      H=\frac{1}{2}[\pi_{t}^{2}+(\nabla\phi_{t})^{2}+\mu^{2}\phi_{t}^{2}]+
      V(\phi_{t}).
\end{equation}

\smallskip
1. \emph{Operator formulation.}

The operator formulation was introduced by
W.~Heisenberg and W.~Pauli in 1929. They defined
quantum fields as unbounded selfadjoint operator-valued fields
$\hat{\phi}_{t}$ on the  space
$\mathbf{R}^{d}$ that  satisfy (in the  modernized form)
   the canonical commutation   relations
\begin{equation}
[\hat{\phi}_{t}, \hat{\pi}_{t}] =i\int\!dx\:\phi_{t}(x)\pi_{t}(x),
\end{equation}
and  the  classical partial differential Euler-Lagrange  equations on
$\mathbf{R}^{d}\times\mathbf{R}$ associated with
the Lagrangean $L$
\begin{equation}
\frac{d}{dt}\frac{\partial L}{\partial\dot{\phi}}
+ \nabla \frac{\partial L}{\partial\nabla\phi}-
\frac{\partial L}{\partial\phi}=0.
\end{equation}
     In our example
\begin{equation}
\partial_{t}^{2}\hat{\phi}_{t}-\nabla^{2}\hat{\phi}_{t}+
\mu^{2}\hat{\phi}_{t}+V'(\hat{\phi}_{t})=0.
\end{equation}
If \emph{non-linear},  the Euler-Lagrange equations  are ill suited
for  operator valued solutions. Yet
they can be solved  explicitly   when $V=0$ (i.e. for free  fields).
     For non-zero $V$  physicists  approximate solutions
     using the perturbation series expansion or lattice
discretization. The numerical results are in amazing  agreement
     with experiments.
However both approximations diverge in the limit.
It is suggestive that   relativistic theory is impossible if  $V\neq 0$
and $d>3$  (cf.\cite{Bau88}).

2. \emph{Functional differential formulation.}

The functional differential formulation was introduced by S. Tomonaga in 1946
     and by J. Schwinger in 1951. We present it in a non-relativistic   form.

The quantum states $\Psi_{t}(\phi)$
are time dependent functionals of classical fields satisfying
the  \emph{linear} functional differential Schr\"{o}dinger equation (in
appropriate units)
\begin{equation}
\frac{\partial }{\partial t}\Psi_{t}
+iH(\hat{\phi}, \hat{\pi})\Psi_{t}=0.
\end{equation}

     \smallskip
For comparison, in quantum mechanics the quantum states are functions
$\phi_{t}(q)$
of the position vectors $q\in \mathbf{R}^{d}$. The conjugate momentum
vectors are $p\in\mathbf{R}^{d}$, so that
the   Schr\"{o}dinger equation is
\begin{equation}
\frac{\partial }{\partial t} \psi_{t}(q)
+iH(\hat{q}, \hat{p}) \psi_{t}(q)=0,
\end{equation}
where   $\hat{q}$ and $\hat{p}$
satisfy the canonical commutation relations
$[\hat{q},\hat{p}]=i(q\!\cdot\!p)$.
Because canonically conjugate  $\hat{q}$ and $\hat{p}$ do not commute,
the meaning of the Hamiltonian operator $H(\hat{\phi}, \hat{\pi})$
is ambiguous.

In 1927 H.~Weyl proposed a definition of $H(\hat{q}, \hat{p})$
     based on the group-theoretical
approach to canonical commutation relations
(cf. \cite{Wey50}). That was the first appearance of a
pseudodifferential operator.
A formal calculus of Weyl operators was developed by
J. Moyal \cite{Moy49}. In the 60's
mathematical physicists found many other heuristic  symbolic calculi
of Weyl operators, the most general by C. Agarwal an E. Wolf \cite{AW70}.
Rigorous mathematics was established  by F.~Berezin \cite{Ber71}
for wick and  antiwick symbols, by L. H\"{o}rmander
\cite{Hor79} for  the Weyl-Moyal calculus, and in  \cite{Dyn98}
for the   Agarwal-Wolf calculus. (See also Chapter 4 of \cite{Shu01}.)

\smallskip
Various  versions of \emph{infinite-dimensional
pseudodifferential operators} have been  introduced
by M. Vishik (cf. \cite{Vis71}),  O. Smolyanov and A. Khrennikov
\cite{SK87} (cf. also \cite{Khr89}), and in a greater  generality
by  P. Kr\'{e}e (cf. \cite{Kre77}).
However their symbolic calculus is not sufficient for a full-fledged
theory of ellipticity, with the notable exception of B.Lascar work
\cite{Las77}. The latter is  an infinite-dimensional version  of
Berezin theory \cite{Ber71}. Significantly, the
Berezin-Lascar definition is not based on  Fourier transform.

3. \emph{Functional integral formulation.}

The functional integral formulation of quantum mechanics was introduced
by R. Feynman in 1942 and of quantum field theory in 1950.

Formally independent of the  differential Schr\"{o}dinger equation
the formulation  starts with the  solution  $\Psi_{t}$ of (1.6) in the
functional integral form:
\begin{equation}
\Phi_{t}(\phi_{t})=\int\! d\phi_{0}\: \langle \phi_{t} |\phi_{0}\rangle
\Phi_{0}(\phi_{0}),
\end{equation}
where   the integral kernel $\langle \phi_{t} |\phi_{0}\rangle$
is considered to be the \emph{quantum propagator of  classical fields}
from $\phi_{0}$ to  $\phi_{t}$.

In quantum mechanics  the integral  is finite-dimensional:
\begin{equation}
\psi_{t}(q_{t})=\int\!dq_{0}\: \langle q_{t} |q_{0}\rangle
\psi_{0}(q_{0}),
\end{equation}
so that $\langle q_{t} |q_{0}\rangle$ is a Green function for
the Cauchy problem for the Schr\"{o}dinger equation. Actually
the Green function is an oscillatory distribution. Therefore the
integral (1.9) has to be understood in a distributional sense.

Starting with a  Dirac idea, Feynman  came up with the remarkable
     representation of  the  quantum propagator
     $\langle \phi_{t} |\phi_{0}\rangle$
as a ``sum over classical  histories'' $\phi_{\tau}, 0\leq
\tau\leq t,$  on the  space $\mathbf{R}^{d}$:
\begin{equation}
\langle \phi_{t} |\phi_{0}\rangle =
\sum_{\phi_{0}}^{\phi_{t}} \exp i\int_{0}^{t}\!d\tau
\int_{\mathbf{R^{d}}}dx\,L(\phi_{\tau},\nabla\phi_{\tau},\dot{\phi_{\tau}}).
\end{equation}
Due to ``interference'' of the oscillating exponentials, the
     relevant histories are in the vicinity of the classical one, i.e.,
the solution of the   Euler-Lagrange equation. So, in principle,
this ``sum'' would be  semi-classical.

In the first publication \cite{Fey48} Feynman gave only a  quantum mechanical
version of such ``sum'' based on  a heuristic physical picture of multiple
scattering of   free particles on the potential.
His  derivation was based on
     a semi-classical postulate for short time quantum propagators.

A shorthand notation for the ``sum'' is the  \emph{Lagrangean Feynman
integral}
\begin{equation}
\int_{\phi_{0}}^{\phi_{t}}\! \mathcal{D}[\phi]\:
     e^{i\int_{0}^{t}\!d\tau
\int_{\mathbf{R^{d}}}dx\,L(\phi_{\tau},\nabla\phi_{\tau},\dot{\phi_{\tau}})}.
\end{equation}
This compact notation suggests  iterated
integration,  integration by
parts,  substitution rule, WKB approximation, Gaussian integrals, etc.

Most of  mathematical  pursuit of a rigorous Lagrangean Feynman integral
is within the frame of the quantum mechanical Schr\"{o}dinger equation
(cf.  \cite{JL00}), usually for  Hamiltonians of the form
``Minus Laplacian plus scalar potential''.

\smallskip
Another form of the Feynman integral is the \emph{ Hamiltonian Feynman
integral}
\begin{equation}
\int_{\phi_{0}}^{\phi_{t}}\!\mathcal{D}[\phi_{\tau},\pi_{\tau}]
\:e^{i\int_{0}^{t}\!d\tau
\int_{\mathbf{R}^{d}}dx\, [\pi_{\tau}\dot{\phi}_{\tau}-
H(\phi_{\tau},\pi_{\tau})]}.
\end{equation}
The Hamiltonian version for quantum
field theory was proposed by R.~Feynman \cite{Fey51}  in 1951.
     W.~Tobocman \cite{Toc56} derived in 1956 its quantum mechanical version
modifying  Feynman's heuristic  semi-classical postulate.

In 1960 J. Klauder \cite{Kla60} introduced the formal
   \emph{Hamiltonian Feynman integral over the coherent state histories}
   for the  propagator of the coherent states,in the neat agreement with
Feynman equation (1.10)  since the quantum coherent states
     are labeled by the classical fields.

     In 1984 J. Klauder and I. Daubechies \cite{KD84}
gave a rigorous definition  of the Klauder-Feynman  integral in
quantum mechanics with the Hamiltonian function  $H$  replaced
by the antinormal (i.e., antiwick) symbol of the Hamiltonian operator.

Their definition,  unlike the original sequential  Feynman's construction,
is a limit of Wiener
integrals over the \emph{phase space}.  It is
applicable, e.g., to  selfadjoint differential operators.
(However the non-trivial selfadjointness property is only postulated.)

A sequential Hamiltonian  quantum-mechanical Feynman type  integral
is proposed in \cite{Dyn98}. There the convergence proof is based on
the pseudodifferential calculus.

\smallskip
A rigorous construction of Lagrangean Feynman type
integral for self-interacting boson fields can be found in
   \cite{GJ87}, mostly for  $d<3$.

    \subsection{Content  of the paper}
     Section 2  sketches  an infinite-dimensional pseudodifferential
     symbolic calculus
   and an associated ellipticity theory for functional quasi differential
   operators. ( The complete details are  in a forthcoming paper.)

The exposition is based on B. Lascar's infinite-dimensional extension
\cite{Las77}
to functional differential operators   of   Berezin's theory \cite{Ber71}
of \emph{antiwick symbols} on $\mathbf{R}^{n}$.

Our theory is for quasi differential operators
in all Fock representations of the bosonic canonical commutation relations,
in  particular in the  Cook-Fock,   Friedrichs-Fock, and
Bargmann-Fock models.
Appropriately we do not use the infinite-dimensional Fourier transform.

At the end of the section we give rather general sufficient conditions
for essential selfadjointness of functional quasi differential
   operators.

\smallskip
     The  theorem 3.1 of  Section 3  presents  a rigorous Hamiltonian
    Klauder-Feynman    sequential integral for the quantum propagators
of classical
     fields generated by  elliptic functional quasi differential
     operators. It is based on a suitable version of the Feynman-Tobocman
    semi-classical postulate.  Being a limit of multiple functional integrals
     over the phase space, the integral satisfies   standard rules of
     the elementary integral calculus including the substitution rule
     for linear canonical transformations. The convergence proof is based on
     the antiwick symbolic calculus of Section 2.

\subsection{Notational convention}
     Throughout the paper the expression  $A\prec B$ for variable $A$ and $B$
     means that there exists a constant $C>0$ such that $|A|\leq C|B|$.

\section { Quasi differential  operators over Hilbert spaces}
\subsection{ Hilbert spaces with conjugation}

Let $\mathcal{H}$ be a complex separable Hilbert space  with
a \emph{conjugation} $\psi\mapsto\psi^{*}$,
     an antilinear norm-preserving involution on $\mathcal{H}$.

As in \cite{Ber66}, $\psi_{1}^{*}\psi_{2}$ denotes the Hermitean
inner product of $\psi_{1}$ and $\psi_{2}$.

\smallskip
The \emph{real part  of} $\mathcal{H}$ is the real Hilbert subspace
$\Re\mathcal{H}=\{\phi\in\mathcal{H}: \ \phi^{*}=\phi\}$ with the
inner product $\phi_{1} \phi_{2}$. In quantum  theory $\Re\mathcal{H}$
   represents the configuration space.

\smallskip
As a real Hilbert space,  $\mathcal{H}$ is a symplectic vector space
with the symplectic form $\Im(\alpha^{*}\beta)$, the imaginary part
of the Hermitean product. As such, $\mathcal{H}$ is the phase  space of
quantum  field theory.

\smallskip
The \emph{antidual} $\mathcal{H}^{*}$ is the Hilbert
space of continuous antilinear functionals on $\mathcal{H}$.
     By Riesz theorem, they may be represented as  $\psi^{*}$.

The Hilbert space $\mathcal{H}^{*}\times\mathcal{H}$  carries
the  conjugation $(\alpha^{*},\beta)=(\beta^{*},\alpha)$. The
corresponding  real part $\mathcal{R}$
is the \emph{antidiagonal} $\{(\psi^{*},\psi):\psi\in\mathcal{H}\}$.
The isometry $\psi\mapsto (1/\sqrt{2})(\psi^{*},\psi)$ is
a representation of $\mathcal{H}$ as a real Hilbert space.

\subsection{ Fock quantization}

A \emph{Fock quantization} is a pair
of linear representations of $\mathcal{H}$ by closable operators
     $q^{+}(\psi)$ and  $q^{-}(\psi)$
on a dense subspace $\mathcal{F}_{0}$ of another Hilbert space
$\mathcal{F}$ such that:
\begin{enumerate}
\item   On $\mathcal{F}_{0}$ the Hermitean adjoint of
$q^{-}(\psi^{*})$  is $q^{+}(\psi)$.
\item   The commutators satisfy the \emph{Fock commutation relations}
\begin{equation}
\begin{split}
[q^{-}(\alpha^{*}), q^{+}(\beta)]
&=\alpha^{*}\beta, \\
[q^{+}(\alpha), q^{+}(\beta)]=\
     &0 =[q^{-}(\alpha), q^{-}(\beta)].
\end{split}
\end{equation}
\item There is a distinguished unit element
$\Omega_{0}\in\mathcal{F}_{0}$ such that  $q^{-}(\psi)\Omega_{0}=0$
for all $\psi$.
\item The subspace $\mathcal{F}_{0}$  is the
linear span of $q^{+}(\psi)^{n}\Omega_{0}, \
\psi\in\mathcal{H}, \ n\in\mathbf{N}$.
\end{enumerate}
Note that $\mathcal{F}_{0}$ is invariant for both $q^{+}(\psi)$ and
$q^{-}(\psi)$.

The quadruple $(\mathcal{F}, \ \Omega_{0}, \ q^{+}, q^{-})$ is defined
by these axioms uniquely up to unitary  equivalence.

\smallskip
The Hilbert space $\mathcal{F}$ is called a \emph{Fock space} over
$\mathcal{H}$. Its elements $\Psi$ are  \emph{quantum states}.
The Hermitean
inner product of $\Xi$ and $\Upsilon$ in $\mathcal{F}$
is denoted $\langle\Xi| \Upsilon\rangle$ with Dirac assumption
that it is  antilinear on the left and linear on right.
The   state  $\Omega_{0}$ is called the \emph{vacuum state}.
The operators $q^{+}(\psi)$ are \emph{creation operators}
and $q^{-}(\psi)$ are \emph{annihilation operators}.

\subsection{Functional integral}
An increasing sequence of $n$-dimensional subspaces  $\mathcal{H}^{(n)}$
in $\mathcal{H}$ is said to be
\emph{complete} if their union  is dense in $\mathcal{H}$.

Let $\Psi=\Psi(\phi)$ be a (non-linear) functional on
$\Re\mathcal{H}$,
and $\Psi^{(n)}=\Psi^{(n)}(\phi^{(n)})$ be the restrictions of $\Psi$ on
$\Re\mathcal{H}^{(n)}$.

     The  \emph{functional integral} of $\Psi$ over $\Re\mathcal{H}$
is defined (cf. \cite{Ber66}) as the limit of the Lebesgue integrals
over $\Re\mathcal{H}^{(n)}$

\begin{equation}
\int\!d\phi\: \Psi(\phi)=
\lim_{n\rightarrow\infty}\frac{1}{(\pi)^{n/2}}
\int\!d\phi^{(n)} \Psi^{(n)}(\phi^{(n)}),
\end{equation}
if  the limit is the same for all
complete sequences of $n$-dimensional subspaces in $\mathcal{H}$.
The finite-dimensional renormalizations are chosen so that the
     Gaussian functional integral
\begin{equation}
\int \!d\phi\:e^{-\phi\phi}=1.
\end{equation}
The functional  integral is  a positive linear functional on the space
of integrable functionals. The integral over a product Hilbert space
is equal to the iterated  integrals.
The integral is invariant under translations by all elements of
$\Re\mathcal{H}$.

The integration by parts involves the  derivatives in the
   directions of $\alpha\in \Re\mathcal{H}$
\begin{equation}
\int\!d\phi \: \Xi\frac{\partial \Upsilon}{\partial \alpha} =
- \int\!d\phi \: \Upsilon\frac{\partial \Xi}{\partial \alpha}
\end{equation}
   if   $\Xi\Upsilon\rightarrow 0 $ as
     $\phi\phi\rightarrow\infty$ and both integrals exist.

\smallskip
For  functionals  $\Psi=\Psi(\psi^{*},\psi)$ on
$\mathcal{R}$
the \emph{functional integral}
is   the  limit of the  integrals over $\mathcal{R}^{(n)}=
\Re(\mathcal{H}^{*(n)}\times\mathcal{H}^{(n)})$.
\begin{equation}
\int\!d\psi^{*}\!d\psi\: \Psi(\psi^{*},\psi)=
\lim_{n\rightarrow\infty}\frac{1}{(2i\pi)^{n}}\int\!d\psi^{*(n)}d\psi^{(n)}\:
\Psi^{(n)}(\psi^{*(n)},\psi^{(n)})
\end{equation}

\newpage

\subsection{Examples of Fock quantizations}
\subsubsection{ Cook tensor quantization}

     The Cook   space $\mathcal{F}_{0}$ is the  space of symmetric tensors
on  $\mathcal{H}$ (cf. \cite{Coo53} and \cite{RS72},  Section 2.4).
The Fock space is its  Hilbert space completion.
The creation operators $q^{+}(\psi)$   are symmetrized exterior products
     with $\psi$ on the right. The vacuum state is $\Omega_{0}=1$.
     The annihilation operators $q^{-}(\psi)$ are symmetrized
     contractions  with $\psi$ on the left.

\subsubsection{ Friedrichs functional quantization}

       The Friedrichs  Fock space is  the Gaussian Hilbert space
$\mathcal{L}^{2}(\Re\mathcal{H}, e^{-\phi\phi}d\phi)$
     (cf. \cite{Fri53}).
      The  space $\mathcal{F}_{0}$ is the  space
     of continuous polynomials on  $\Re\mathcal{H}$. The vacuum state is
      $\Omega_{0}=e^{-\phi\phi/2}$.  For $\phi\in\Re\mathcal{H}$, the
annihilation
and  creation operators are      the  functional
differential operators $q^{-}(\phi)= \partial/\partial\phi-\phi$ and
$q^{+}(\phi)= \partial/\partial \phi + \phi$. To  all
     $\psi\in\mathcal{H}$  they are extended by the complex linearity.

\subsubsection{Bargmann functional quantization}

     The Bargmann space $\mathcal{F}_{0}$ is the  space of continuous analytic
     polynomials $\Pi=\Pi(\psi^{*})$ on  $\mathcal{H}^{*}$.
     The annihilation operators $q^{-}(\psi)$ are
     the directional derivatives  $\partial/\partial\psi^{*}$.
     The creation operators $q^{+}(\psi)$ are multiplications with the
     linear forms
$\psi$ on $\mathcal{H}^{*}$.  The vacuum  state is $\Omega_{0}=1$.

     The Bargmann  Fock space  $\mathcal{B}$ is the closed subspace
of continuous antiholomorhic functionals $\Psi(\psi^{*})$ in
     $\mathcal{L}^{2}(\mathcal{H}^{*}\times\mathcal{H}, e^{-\psi^{*}\psi}
     d\psi^{*} d\psi)$
(cf. \cite{Bar62} and \cite{Ber66}).

\subsubsection{Geometric quantizations}
There is a functional Fock quantization
associated with every Lagrangean subspace of the symplectic space
$\mathcal{H}^{*}\times\mathcal{H}$ (cf. \cite{Woo92}). E.g.,
the Friedrichs quantization  is associated with $\mathcal{R}$
     and  the Bargmann quantization  with $\mathcal{H}^{*}\times\{0\}$.

\subsection{Hermite polynomials and coherent states}
The quantum  states
     \begin{equation}
\prod_{j=1}^{n}q^{+}(\alpha_{j})\Omega_{0}\in \mathcal{F}_{0}, \
\alpha_{1},\ldots \alpha_{n}\in \mathcal{H},
\end{equation}
are called Hermite polynomials of order $n$.

E.g., in the Bargmann space they are
$\prod_{j=1}^{n}(\psi^{*}\alpha_{j})$.

By polarization, the  Hermite polynomials  are linear
combinations of the Hermite monomials $q^{+}(\alpha)^{n}\Omega_{0}$.

We have (cf. (6.3.12) of \cite{GJ87})
\begin{equation}
\langle q^{+}(\alpha)^{m}\Omega_{0}|q^{+}(\beta)^{n}\Omega_{0}\rangle=
\delta_{mn}n!(\alpha^{*}\beta)^{n}.
\end{equation}
This shows that the correspondence between $\alpha$ and
$q^{+}(\alpha)^{n}\Omega_{0}$ is one to one.
It follows also that if $\{\epsilon_{j}\}\subset\mathcal{H}_{+}$ is an
orthonormal basis in $\mathcal{H}$, then
the Hermite monomials $q^{+}(\epsilon_{j})^{n}\Omega_{0}$
form an orthogonal basis in  $\mathcal{F}$ with
     $|\!|q^{+}(\epsilon_{j})^{n}\Omega_{0}|\!|
     =\sqrt{n!}$.

\smallskip
The  quantum states
\begin{equation}
\Omega_{\alpha} =
\sum_{n=1}^{\infty}\frac{1}{n!} q^{+}(\alpha)^{n}\Omega_{0}, \
\alpha\in\mathcal{H},
\end{equation}
are called (non-normalized) \emph{ coherent states}.

In the Bargmann space,
$\Omega_{\alpha}(\psi^{*})=e^{\psi^{*}\alpha}$.

In view of (2.7), we have
\begin{equation}
\langle \Omega_{\alpha} |\Omega_{\beta}\rangle=
e^{\alpha^{*}\beta}.
\end{equation}
This equality implies that $\Omega_{\alpha}$ belong to $\mathcal{F}$ and
that the correspondence between $\alpha$ and $\Omega_{\alpha}$ is one to one.

We have, by (2.7),
\begin{equation}
q^{+}(\beta)\Omega_{\alpha}(\psi^{*})=(\beta\psi^{*})\Omega_{\alpha}(\psi^{*}),
\ q^{-}(\beta^{*})\Omega_{\alpha}(\psi^{*})
=(\beta^{*}\alpha)\Omega_{\alpha}(\psi^{*}).
\end{equation}
The coherent states form an overcomplete orthogonal basis, i.e, if
      $\Xi, \Upsilon\in\mathcal{F}$ then we have the analogue of the Plancherel
      equality
\begin{equation}
     \langle\Xi |\Upsilon\rangle = \int\!d\psi^{*}d\psi
     \:e^{-\psi^{*}\psi}
\: \langle\Xi| \Omega_{\psi}\rangle\langle\Omega_{\psi}|\Upsilon\rangle.
\end{equation}

     To verify the equation it suffices to check it on the total set of
     Hermite monomials. For them the equation holds because of (2.7).

In other terms, every $\Psi\in\mathcal{F}$ has an analogue of the Fourier
expansion
\begin{equation}
\langle \Omega_{\alpha}|\Psi\rangle = \int\!d\psi^{*}d\psi
     \:e^{-\psi^{*}\psi}e^{\alpha^{*}\psi}
\: \langle\Omega_{\psi}|\Psi\rangle.
\end{equation}

\subsection{Sobolev Fock  spaces $\mathcal{F}^{s}$}

Suppose  $h$ is a real  non-negative
selfadjoint  operator in  $\mathcal{H}$ such that  $(1 + h)^{-1}$
     is a  Hilbert-Schmidt operator.
Let
\[
|\!|\psi|\!|_{+}^{2}=\psi^{*}(1 + h)\psi, \
|\!|\psi|\!|_{-}^{2}=\psi^{*}(1 + h)^{-1}\psi.
\]
Then we have the Hilbert space $\mathcal{H}_{+}\subset \mathcal{H}$
of all $\psi$ with finite norm $|\!|\psi|\!|_{+}$ and the Hilbert space
$\mathcal{H}_{-}$, that is the completion of $\mathcal{H}$ with respect to
the norm $|\!|\psi|\!|_{-}$. These spaces inherit the conjugation
from  $\mathcal{H}_{-}$. We got the nested triple
\[
\mathcal{H}_{+}\subset\mathcal{H}\subset\mathcal{H}_{-},
\]
where both inclusions are Hilbert-Schmidt operators with dense ranges.

The \emph{ Sobolev space} $\mathcal{B}^{s}, \
s\in\mathbf{R},$ is the Hilbert space of continuous holomorphic functionals
$\Psi(\psi^{*})=\langle\Omega_{\psi}|\Psi\rangle$ on
$\mathcal{H}^{*}$ with finite norm $|\!|\Psi|\!|_{s}$
defined by (cf. \cite{Las77}, Definition 2.2)
\begin{equation}
|\!|\Psi|\!|_{s}^{2}=\int\!d\psi^{*}d\psi\:
e^{-\psi^{*}\psi}(1+|\!|\psi|\!|_{-})^{s}|\Psi(\psi^{*})|^{2}.
\end{equation}
Every $\mathcal{B}^{s}$ contains all continuous holomorphic polynomials on
$\mathcal{H}^{*}$.

The topological intersection $\mathcal{B}^{\infty}=\cap_{s}\mathcal{B}^{s}$
is a Frechet space. Its  topological  dual is
$\mathcal{B}^{-\infty}=\cup_{s}\mathcal{B}^{s}$.

By  definition, the Hilbert space $\mathcal{A}^{s},\
-\infty < s <\infty$,  consists of
the functionals $\Psi(\psi^{*},\psi)$ on $\mathcal{R}$ with finite
norm defined as in (2.13).

\begin{remark}[ Proposition 2.4 of \cite{Las77}]
The operator
\begin{equation}
\Psi(\psi^{*},\psi)\mapsto\int\!d\xi^{*}d\xi\: e^{-\xi^{*}\xi}
e^{\psi^{*}\xi}\Psi(\xi^{*},\xi)
\end{equation}
is the orthogonal projector of  $\mathcal{A}^{s}$ onto
$\mathcal{B}^{s}$.
\end{remark}
Since the Fock quantization is unique up to unitary equivalence,
we have counterparts $\mathcal{F}^{s}$ of $\mathcal{B}^{s}$
in any representation of Fock commutation relations.

\subsection{Functional quasi differential operators}

   Consider an \emph{antiwick monomial operator}
\begin{equation}
Q_{\{\alpha_{j}\}}^{k,l}=\prod_{j=1}^{k}q^{-}(\alpha_{j}^{*})
\prod_{j=k+1}^{k+l}q^{+}(\alpha_{j}),
\end{equation}
where $\alpha_{j}
\in\mathcal{H}_{+}, \ j=1,2,\ldots,k+l$.
Note that this is  a functional differential operator
in any functional  Fock  space   (2.4).

In the Bargmann space $\mathcal{B}$ we have, by 2.4.3 and (2.12),
\begin{equation}
\begin{split}
&Q_{\{\alpha_{j}\}}^{k,l}\Psi(\psi^{*})
= \prod_{j=1}^{k}\partial_{\alpha_{j}^{*}}
[\prod_{j=k+1}^{k+l}(\psi^{*}\alpha_{j})\Psi(\psi^{*})] \\
&=\int\!d\xi^{*}d\xi\: e^{-\xi^{*}\xi}e^{\psi^{*}\xi}
P_{\{\alpha_{j}\}}^{k,l}(\xi^{*},\xi)\Psi(\xi^{*}),
\end{split}
\end{equation}
where
\begin{equation}
P_{\{\alpha_{j}\}}^{k,l}(\xi^{*},\xi)=
\prod_{j=1}^{k}\alpha_{j}^{*}\xi\prod_{j=k+1}^{k+l}\xi^{*}\alpha_{j},
\ \alpha_{j}\in \mathcal{H}_{+},
\end{equation}
is the Hermite polynomial on $\mathcal{R}$ that is continuous in
the norm $|\!|\cdot|\!|_{-}$.
In view of (2.14), the monomial operator $Q_{\{\alpha_{j}\}}^{k,l}$
is  bounded  from $\mathcal{F}^{s}$ to $\mathcal{F}^{s-k-l}$  for all $s$.

The same holds for the  general  \emph{antiwick operator
$Q(A)$ of order} $\rho\in\mathbf{R}$
\begin{equation}
Q(A)\Psi(\psi^{*})=
\int\!d\xi^{*}d\xi\: e^{-\xi^{*}\xi}e^{\psi^{*}\xi}
A(\xi^{*},\xi)\Psi(\xi^{*}),
\end{equation}
where $A(\xi^{*},\xi)\prec(1+ |\!|\xi|\!|_{-})^{\rho}$ and is continuous
in the norm $|\!|\cdot|\!|_{-}$.

Let $\{e_{j}\}\subset\mathcal{H}_{+}$ be an orthonormal basis in
$\mathcal{H}$. Then   we have the Hermite  orthogonal expansion of $A$
with constant coefficients $c_{\{j_{1},\ldots,j_{k+l}\}}^{k,l}$
\begin{equation}
A=\sum c_{\{j_{1},\ldots,j_{k+l}\}}^{k,l}
P_{\{e_{j_1},\dots, e_{j_{k+l}}\}}^{k,l},
\ \sum (k+l)!
|c_{\{j_{1},\ldots,j_{k+l}\}}^{k,l}|^{2}<\infty.
\end{equation}
Now (2.18) and (2.19) imply the  expansion of the antiwick
   operator $Q(A)$ into antiwick monomial operators:
\begin{equation}
Q(A)=\sum c_{\{j_{1}, \ \ldots,j_{k+l}\}}^{k,l}
Q(P_{\{e_{j_1},\dots, e_{j_{k+l}}\}}^{k,l}),
\ \sum (k+l)!|c_{\{j_{1}, \ \ldots,j_{k+l}\}}^{k,l}|^{2}<\infty.
\end{equation}

As in \cite{Ber71},  the correspondence between $A$ and $Q(A)$ is one
to one. Accordingly, $A$ is called the \emph{antiwick symbol} of
$Q(A)$.
In view of (2.14)  the operator norm of $Q(A)$ in $\mathcal{F}$
\begin{equation}
|\!|Q(A)|\!|\leq\sup|A(\xi^{*},\xi)|
\end{equation}
and, if $A(\xi^{*},\xi)$ is bounded from below  by a constant $c$, then
\begin{equation}
\langle \Psi|Q(A)|\Psi\rangle \geq c\langle \Psi|\Psi\rangle.
\end{equation}
By the definition (2.13) and Remark 2.1,
it follows that  $Q(A)$ of order $\rho$ is a bounded operator from
$\mathcal{F}^{s}$ to $\mathcal{F}^{s-\rho}$ with the operator
norm
\begin{equation}
|\!|Q(A)|\!|_{s,s-\rho}\prec \sup|A(\psi^{*},\psi)|/
(1+|\!|\psi|\!|_{-})^{\rho}.
\end{equation}
     In consequence, $Q(A)$ are continuous
operators in $\mathcal{F}^{\infty}$ and $\mathcal{F}^{-\infty}$.

In view of (2.9) and (2.11), the \emph{coherent states matrix
element of} $Q(A)$ is
\begin{equation}
\langle \Omega_{\alpha}|Q(A)|\Omega_{\beta}\rangle=
\int\!d\xi^{*}\!d\xi\:e^{-\xi^{*}\xi}A(\xi^{*},\xi)
e^{\alpha^{*}\xi+\xi^{*}\beta}.
\end{equation}
This is an entire function in $(\alpha^{*},\beta)$ on
$\mathcal{H}^{*}\times\mathcal{H}$. As such it is uniquely defined  by
its restriction $\langle \Omega_{\psi}|Q(A)|\Omega_{\psi}\rangle$
to the antidiagonal $\mathcal{R}$.

The \emph{wick (i.e., normal) symbol}, $A^{[\nu]}(\psi^{*},\psi)$ of $Q(A)$,
is defined by the coherent matrix element (cf. \cite{Ber71}) as
\begin{equation}
A^{[\nu]}(\psi^{*},\psi)=
\langle\Omega_{\psi}|Q(A)|\Omega_{\psi}\rangle e^{-\psi^{*}\psi}.
\end{equation}
Actually  the right side of (2.25) defines the wick symbol
$A(Q)^{[\nu]}(\psi^{*},\psi)$ for \emph{any} continuous operator $Q$ from
$\mathcal{F}^{\infty}$ to $\mathcal{F}^{-\infty}$ even when
$Q$ is not an antiwick operator.

\smallskip
An antiwick operator $Q(A)$ is called \emph{quasi differential
of order} $\rho\in\mathbf{R}$ if
its antiwick symbol   $A(\psi^{*},\psi)$  is
a \emph{quasi polynomial of order} $\rho$, i.e.,
     $A(\psi^{*},\psi)$ is $C^{\infty}$ Frechet differentiable,
       the $j$-th differentials  $D^{j}A(\psi^{*},\psi; d\psi^{*},d\psi)$
       are continuous polynomials in $(d\psi^{*},d\psi)$ with respect
      to the weaker  norm $|\!|d\psi|\!|_{-}$,  and
      for every $j$ the norm
      \begin{equation}
      |\!|\!|D^{j}A(\psi^{*},\psi)|\!|\!|_{+}=\sup_{0\neq
      d\psi\in\mathcal{H}_{-}}
     |D^{j}A(\psi^{*},\psi; d\psi^{*},d\psi)|/ |\!|d\psi|\!|_{-}^{j}
     \end{equation}
     satisfies
\begin{equation}
|\!|\!|D^{j}A(\psi^{*},\psi)|\!|\!|_{+}
\prec (1+|\!|\psi|\!|_{-})^{\rho-j}.
\end{equation}
In particular, $A(\psi^{*},\psi)$ is continuous with respect to the
     norm $|\!|\cdot |\!|_{-}$ and
\begin{equation}
A(\psi^{*},\psi) \prec (1+|\!|\psi |\!|_{-})^{\rho}.
\end{equation}

The linear spaces of the quasi polynomials of the order $\rho$ is denoted
as $\mathcal{Q}^{\rho}$, their intersection as
$\mathcal{Q}^{-\infty}$, and their union as $\mathcal{Q}^{\infty}$.

\subsection{Agarwal-Wolf symbols}
Quasi differential operators $Q(A)$ over $\mathbf{R}^{n}$ are
pseudodifferential operators. However, as   in \cite{Ber71} and \cite{Las77},
they  are defined without the Fourier transform. Yet starting from
the antiwick symbol one may define all
its Agarwal-Wolf pseudodifferential symbols including the most common.

For the starter, consider a formal power series
on $\mathcal{H}^{*}\times\mathcal{H}$
\begin{equation}
\omega(\xi^{*},\xi) = 1+\sum_{n=1}^{\infty}\omega_{n}(\xi^{*},\xi),
\end{equation}
where $\omega_{n}$ are continuous polynomials of degree $n$ on
$\mathcal{H}_{-}^{*}\times\mathcal{H}_{-}$.

The \emph{formal $\omega$-symbol of the  operator} $Q(A)$ is defined by
\begin{equation}
A^{\omega}(\psi^{*},\psi)=
\omega(\partial_{\psi^{*}},-\partial_{\psi})A(\psi^{*},\psi) = 1+
\sum_{n=1}^{\infty}\omega_{n}(\partial_{\psi^{*}},-\partial_{\psi})
A(\psi^{*},\psi).
\end{equation}
Applying the classical Borel-H\"{o}rmander construction (cf.
in \cite{Shu01},  Proposition 3.5) to $A\in\mathcal{Q}^{\rho}$
we get a quasi polynomial $A^{[\omega]}\in\mathcal{Q}^{\rho}$
     such that for  $N=1,2,\ldots$
\begin{equation}
A^{[\omega]}-\sum_{n<N}\omega_{n}(\partial_{\psi^{*}},-\partial_{\psi})
A(\psi^{*},\psi)\in\mathcal{Q}^{\rho-N}.
\end{equation}
The differences between such $A^{[\omega]}$ belong to
$\mathcal{Q}^{\infty}$.

     In particular, the formal antiwick symbol $A^{1}$  corresponds to
     $\omega=1$.
     Among other  formal symbols  we have  (cf.\cite{AW70})
\begin{enumerate}
\item \emph{Wick (or normal)  symbol  $A^{\nu}$ with
     $\omega(\psi^{*},\psi) = e^{-\psi^{*}\psi}$}.
\item \emph{Weyl  symbol $A^{\gamma}$ with
$\omega(\psi^{*},\psi) = e^{-\frac{1}{2}\psi^{*}\psi}$}.
\item \emph{Left  symbol $A^{\lambda}$ with
     $\omega(\psi^{*},\psi) =
e^{-\frac{1}{4}(\psi^{*}\psi^{*}-\psi\psi)}e^{-\frac{1}{2}\psi^{*}\psi}$.}
\item \emph{Right symbol $A^{\rho}$ with
$\omega(\psi^{*},\psi) =
e^{+\frac{1}{4}(\psi^{*}\psi^{*}-\psi\psi)}e^{-\frac{1}{2}\psi^{*}\psi}$.}
\item \emph{$\sigma$-symbol $A^{\sigma}$ with $\omega(\psi^{*},\psi) =
e^{\sigma\psi^{*}\psi}e^{-\frac{1}{2}\psi^{*}\psi}, \  \sigma\in\mathbf{C}$.}
\item \emph{$\tau$-symbol $A^{\tau}$ with  $\omega (\psi^{*},\psi) =
e^{\tau(\psi^{*}\psi^{*}-\psi\psi)}e^{-\frac{1}{2}\psi^{*}\psi}
, \ \tau\in\mathbf{C}$.}
\end{enumerate}

In mathematical literature , where
$\Re\mathcal{H}=\mathbf{R}^{n}$,  the left and right symbols
are  called \emph{Kohn-Nirenberg symbols}.

For every $\omega$ the symbols $A^{[\omega]}$ satisfy the inequality
(2.27).

\subsection{Symbolic calculus}
Since the quasi differential operators act on $\mathcal{F}^{\infty}$ (and
$\mathcal{F}^{-\infty}$), the  product
     $Q(B)Q(C)$ of two such operators with $B\in\mathcal{Q}^{\rho}$
and  $C\in\mathcal{Q}^{\tau}$ is well defined.
In fact, the product is a quasi differential operator $Q(A)$
with the  formal antiwick symbol $A^{1}$ (cf. Proposition 3.6 of \cite{Las77}
for differential operators)
\begin{equation}
A^{1}(\psi^{*},\psi)=\sum_{n=0}^{\infty}(-1)^{n}(n!)^{-1}
\partial^{n}B(\psi^{*},\psi)\cdot
\partial^{*n}C(\psi^{*},\psi),
\end{equation}
where the ``dot'' product  denotes
the  contraction  of the polynomial analytic and antianalytic differentials:
\begin{equation}
\int\!d\xi^{*}d\xi\:e^{-\xi^{*}\xi} \partial^{n}B(\psi^{*},\psi)
(\xi^{n})\: \partial^{*n}C(\psi^{*},\psi)(\xi^{*n}).
\end{equation}
Moreover, for every antiwick symbol $A^{[1]}(\psi^{*},\psi)$,
\begin{equation}
A^{[1]}(\psi^{*},\psi)- \sum_{n=0}^{N}(-1)^{n}(n!)^{-1}
\partial_{\psi}^{n}B(\psi^{*},\psi)\cdot
\partial_{\psi^{*}}^{n}C(\psi^{*},\psi)\in
\mathcal{Q}^{\rho + \tau-N}.
\end{equation}

\begin{remark}
Using the relations between Agarwal-Wolf symbols and the antiwick
symbols it is possible to derive the asymptotic expansions of
the  Agarwal-Wolf symbols
for the products of functional pseudodifferential operators.
\end{remark}
\begin{remark}
An  operator $Q(A)$ is symmetric if and only if $A$ is real.
\end{remark}

Using the Plancherel identity (2.11) one gets the coherent state matrix
element of the product $Q(B)Q(C)$
\[
\begin{split}
      &\langle \Omega_{\alpha}|Q(B)Q(C)|\Omega_{\beta}\rangle
=\int\!d\psi^{*}d\psi\:e^{-\psi^{*}\psi}
\langle\Omega_{\alpha}|Q(B)|\Omega_{\psi}\rangle
\langle\Omega_{\psi}|Q(C)|\Omega_{\alpha}\rangle \\
&=\int\!d\psi^{*}d\psi\:e^{-\psi^{*}\psi} \\
&\times\int\!d\xi_{2}^{*}d\xi_{2}\:e^{-\xi_{2}^{*}\xi_{2}}
e^{\alpha^{*}\xi_{2}}B(\xi_{2}^{*},\xi_{2})
\int\!d\xi_{1}^{*}d\xi_{1}\:e^{-\xi_{1}^{*}\xi_{1}}
     e^{\xi_{1}^{*}\beta}C(\xi_{1}^{*},\xi_{1}) \\
&=\int\!d\xi_{2}^{*}d\xi_{2}d\xi_{1}^{*}d\xi_{1}\:e^{-\xi_{2}^{*}\xi_{2}}
e^{-\xi_{1}^{*}\xi_{1}}e^{\alpha^{*}\xi_{2}} e^{\xi_{1}^{*}\beta}
e^{\xi_{2}^{*}\xi_{1}}
B(\xi_{2}^{*},\xi_{2})C(\xi_{1}^{*},\xi_{1}),
\end{split}
     \]
     so that the coherent state kernel of the product $Q(B)Q(C)$
\begin{equation}
\begin{split}
&\langle\Omega_{\alpha}|Q(B)Q(C)|\Omega_{\beta}\rangle
e^{-\beta^{*}\beta} \\
&=\int\!d\xi_{2}^{*}d\xi_{2}d\xi_{1}^{*}d\xi_{1}\:
B(\xi_{2}^{*},\xi_{2})C(\xi_{1}^{*},\xi_{1})
e^{(\alpha^{*}-\xi_{2}^{*})\xi_{2}+(\xi_{2}^{*}-\xi_{1}^{*})\xi_{1}+
(\xi_{1}^{*}-\beta^{*})\beta}.
\end{split}
\end{equation}

\subsection{Families of quasi differential operators }
Consider a parametrized family $Q(A_{\epsilon}), \ 0<\epsilon\leq
\epsilon_{0},$
of quasi differential operators in $\mathcal{Q}^{\rho}$ such that for
some   real number $\mu$
\begin{equation}
|\!|\!|D^{k}A_{\epsilon}(\psi^{*},\psi)|\!|\!|_{+}
\prec \epsilon^{\mu+k}(1+|\!|\psi|\!|_{-})^{\rho-k}
\end{equation}
(cf. Section 29  of \cite{Shu01}).

The linear space of  the quasi polynomial families $A_{\epsilon}$,
   satisfying (2.36),
   is denoted  $\mathcal{Q}_{\epsilon}^{\rho,\mu}$.

Similarly to (2.34),
the product $Q(B_{\epsilon})Q(C_{\epsilon})$ of quasi differential
operator families with $B\in\mathcal{Q}_{\epsilon}^{\rho,\mu}$
and $C\in\mathcal{Q}_{\epsilon}^{\tau,\nu}$
is a quasi differential operator
family $Q(A_{\epsilon})$
such that
\begin{equation}
A_{\epsilon}(\psi^{*},\psi)- \sum_{n=0}^{N}(-1)^{n}(n!)^{-1}
\partial^{n}B_{\epsilon}(\psi^{*},\psi)\cdot
\partial^{*n}C_{\epsilon}(\psi^{*},\psi)\in
\mathcal{Q}_{\epsilon}^{\rho+\tau-N,\mu+\nu+N}.
\end{equation}

\subsection{Ellipticity}
A quasi differential  operator $Q(A)$ of a positive  order $\rho$ is called
\emph{elliptic} if:
\begin{enumerate}
\item
There exists positive $\sigma\leq\rho$ such that
for sufficiently large $|\!|\psi |\!|_{-}$ we have
\[
|\!|\psi|\!|_{-}^{\sigma} \prec A(\psi^{*},\psi).
\]
\item
For given non-negative integers $k$ and $l$, if
     $|\!|\psi |\!|_{-}$ is sufficiently large then
\[
|\!|\!|\partial^{*k}\partial^{l}A(\psi^{*},\psi )|\!|\!|_{+}
\prec A(\psi^{*},\psi )
|\!|\psi |\!|_{-}^{-(k+l)}.
\]
\end{enumerate}
The standard application of the symbolic calculus (cf. \cite{Shu01})
provides a quasi differential \emph{parametrix} $Q(P)$ of order
$-\sigma$ such that $Q(A)Q(P)-1\in\mathcal{Q}^{-\infty}$ and
$Q(P)Q(A)-1\in\mathcal{Q}^{-\infty}$.  This entails that the zero set
of an elliptic operator belongs to $\mathcal{F}^{\infty}$.

On $\mathbf{R}^{n}$ the following remark is Shubin's  Theorem 26.2 in
\cite{Shu01}. The proof is the same.
\begin{remark}
Any elliptic quasi differential operator $Q(A)$ with real $A$ is essentially
selfadjoint on $\mathcal{F}_{0}$. Its closure is $Q(A)$ on the
domain $\Psi\in\mathcal{F}: \ Q(A)\Psi\in\mathcal{F}$.
\end{remark}
\begin{proof}
The operator $Q(A)$ is symmetric on $\mathcal{F}_{0}$.

Let $Q(A)_{0}$ be the restriction of $Q(A)$ on $\mathcal{F}_{0}$,
$\overline{Q(A)_{0}}$  its closure, and $Q(A)_{0}^{*}$ its Hermitian
adjoint in $\mathcal{F}$. The domain of $Q(A)_{0}^{*}$ consists of
$\Psi\in\mathcal{F}$ such that $Q(A)\Psi\in\mathcal{F}$, and
$Q(A)_{0}^{*}=Q(A)$ on this domain.

An essential selfadjointness test of $Q(A)_{0}$ is that the zeros of two
operators $Q(A)_{0}^{*}+i$ and $Q(A)_{0}^{*}-i$ belong to the domain
of $\overline{Q(A)_{0}}$ (cf. \cite{Shu01}, Theorem 26.1).

Here both  operators are elliptic quasi differential so  the zero sets
are within  $\mathcal{F}^{\infty}$. By the closed graph theorem,
$Q(A)$ is continuous in $\mathcal{F}^{\infty}$. Since $\mathcal{F}_{0}$
is dense in $\mathcal{F}^{\infty}$, the test verifies the essential
selfadjointness of $Q(A)_{0}$.
\end{proof}

\begin{remark}
If $\mathcal{H}$ is infinite-dimensional then a generic elliptic quasi
differential operator is not Fredholm. Correspondingly, the spectrum of an
elliptic operator is almost never discrete.
\end{remark}

\section{Klauder-Feynman  integral}
Let  a quasi differential  operator $Q(A)$ be  essentially selfadjoint on
$\mathcal{F}_{0}$.  If $\Psi_{t}$ is a solution of the functional
Schr\"{o}dinger equation (1.6) in a Fock space $\mathcal{F}$
\[
\partial_{t}\Psi_{t} + iQ(A)\Psi_{t}=0
\]
then, by the coherent state Plancherel equality (2.11),
\begin{equation}
\langle \Omega_{\alpha}|e^{-iQ(A)t}|\Psi_{0}\rangle
=\int\!d\beta^{*}d\beta\: e^{-\beta^{*}\beta}
\langle \Omega_{\alpha} |e^{-iQ(A)t}|
\Omega_{\beta}\rangle \langle\Omega_{\beta}|\Psi_{0}\rangle,
\end{equation}
so that
$\langle\Omega_{\alpha} |e^{-iQ(A)t}|\Omega_{\beta}\rangle
e^{-\beta^{*}\beta},
     \ \alpha,\beta\in\mathcal{H},$
is the quantum propagator of the coherent states  (cf. (1.8)).

\begin{theorem}
Suppose  a quasi differential operator $Q(A)$
is elliptic with a real antiwick symbol $A$.
( By Remark 2.4, $Q(A)$ is essentially selfadjoint on  $\mathcal{F}_{0}$.)

Then the coherent state propagator (3.1)
     is equal to the  limit  at $n=\infty$ of the
$n$-multiple functional integrals over the phase space $\mathcal{H}$
\begin{equation}
\int\prod_{j=1}^{n}\!d\psi_{j}^{*}d\psi_{j}\:
\exp\sum_{j=0}^{n}[(\psi_{j+1}-\psi_{j})^{*}\psi_{j} -
iA(\psi_{j}^{*},\psi_{j})t/n],
\end{equation}
where $\psi_{n+1}= \alpha,\ \psi_{0} = \beta$.
\end{theorem}
\begin{proof}

\emph{Step} 1.

By ellipticity of $Q(A)$,  the quasi polynomial family
\begin{equation}
A_{t/n}=(1+iAt/n)^{-1}, \ n=1,2,\ldots,
\end{equation}
parametrized by $\epsilon = t/n$, belongs
to $\mathcal{Q}_{t/n}^{0,0}$.

Applying  (2.37)
to the quasi differential operator  families $1+iQ(A)t/n$ and
$Q(A_{t/n})$,
   we get $(1+iQ(A)t/n)Q(A_{t/n})=Q(D_{t/n})$, where
\[
\begin{split}
&D_{t/n}(\psi^{*},\psi)=
(1+iA(\psi^{*},\psi)t/n)(1+iA(\psi^{*},\psi)t/n)^{-1}\\
&+\partial(1+iA(\psi^{*},
\psi)t/n)\cdot\partial^{*}(1+iA(\psi^{*},\psi)t/n)^{-1}
+(t/n)^{2}R_{t/n}\\
&=1+(t/n)^{2}S_{t/n},
\end{split}
\]
with the quasi polynomial family $S_{t/n}$ uniformly bounded with
respect to $n$.

By (2.21), the operator norms of the quasi differential operators
$Q(S_{t/n})$ in
$\mathcal{F}$ are  uniformly bounded with respect to $n$ as well.
   Therefore
\begin{equation}
|\!|Q(A_{t/n})-(1+iQ(A)t/n)^{-1}|\!| \prec 1/n^{2}.
\end{equation}
We have the telescopic equality
\begin{equation}
\begin{split}
&Q(A_{t/n})^{n}-(1+iQ(A)t/n)^{-n} \\
=& \  \sum_{m=0}^{n-1}Q(A_{t/n})^{m}[(Q(A_{t/n})-(1+iQ(A)t/n)^{-1}]
(1+iQ(A)t/n)^{n-1-m}.
\end{split}
\end{equation}
In view of (2.21), the  operator norm of   $Q(A_{t/n})$ is $\leq 1$,
and the  operator norm of $(1+iQ(A)t/n)^{-1}$ is   $\leq 1$
by the spectral theorem. So (3.5)  entails that, uniformly in $n$,
\begin{equation}
|\!|Q(A_{t/n})^{n}-(1+iQ(A)t/n)^{-n}|\!|\prec n(1/n^{2}).
\end{equation}
Along with the equality
      $\exp[-iQ(A)t]=\lim_{n\rightarrow\infty}(1+iQ(A)t/n)^{-n}$ this
      implies the strong operator limit
\begin{equation}
e^{-iQ(A)t}=\lim_{n\rightarrow\infty}Q(A_{t/n})^{n}.
\end{equation}
\emph{This  equation is a modification of the  semi-classical
Feynman-Tobocman postulate.}

\smallskip
\emph{Step} 2.

The normal symbol of $Q(A_{t/n})$ is given by
\[
\langle\Omega_{\psi_{2}}|Q(A_{t/n})|\Omega_{\psi_{0}}\rangle
e^{-\psi_{0}^{*}\psi_{0}}=
\int\!d\psi_{1}^{*}d\psi_{1}\:
e^{(\psi_{2}-\psi_{1})^{*}\psi_{1}+(\psi_{1}-\psi_{0})^{*}\psi_{0}}
A_{t/n}(\psi^{*},\psi),
\]
so that, by (2.35), the normal symbol of $Q(A_{t/n})^{n}$ is
\begin{equation}
\langle\Omega_{\alpha}|Q(A_{t/n})^{n}|\Omega_{\beta}\rangle
e^{-\beta^{*}\beta}=
\int\!\prod_{j=1}^{n}d\psi_{j}^{*}d\psi_{j}\:A_{t/n}(\psi_{j}^{*},\psi_{j})
e^{\sum_{j=0}^{n}(\psi_{j+1}-\psi_{j})^{*}\psi_{j}}
\end{equation}
with $\psi_{n+1}= \alpha,\  \psi_{0} = \beta$.

\smallskip
\emph{Step} 3.

Consider yet  another  telescopic equality
\begin{equation}
\begin{split}
&Q(A_{t/n})^{n}-Q(e^{-iAt/n})^{n} \\
&=\sum_{m=0}^{n-1}Q(A_{t/n})^{m}[Q(A_{t/n})-Q(e^{-iAt/n})]
Q(e^{-iAt/n})^{n-1-m}.
\end{split}
\end{equation}
The middle factor is the (not a quasi differential)  operator with
the antiwick symbol $A_{t/n}-e^{-iAt/n}$.

Using the second Taylor approximation in the $t$-variable
(with Lagrange  integral remainder) at $t=0$, we get
\[
A_{t/n}-e^{-iAt/n}
=\ (iA/n)^{2}\int_{0}^{t}\!d\tau\:(t-\tau)[-2(1+iA\tau)^{-3}-e^{-iA\tau}]
\prec (At/n)^{2}.
\]
Suppose $\rho$ is the order of $A$.

Then, by  (2.23), $Q(A_{t/n}-e^{-iAt/n})$ is a bounded
     operator from $\mathcal{F}^{\rho}$ to $\mathcal{F}$
   with the operator norm
\begin{equation}
|\!|Q(A_{t/n})-Q(e^{-iAt/n})|\!|_{\rho,0}
\prec (t/n)^{2}\sup |A(\psi^{*},\psi)|/(1+|\!|\psi|\!|_{-})^{\rho}.
\end{equation}
Similarly, the $0$-order operator $Q(e^{-iAt/n})$ is bounded in
   $\mathcal{F}$ with the operator norm $\leq 1$ and the $0$-order operator
$Q(A_{t/n})$ is bounded in $\mathcal{F}^{\rho}$ also with the operator norm
$\leq 1$.

Now   telescopic inequality (3.9) and these norm estimates imply
\begin{equation}
|\!|Q(A_{t/n})^{n}-Q(e^{-iAt/n})|\!|_{\rho,0}
\prec n(t/n)^{2}.
\end{equation}
Therefore, for given $\Omega_{\alpha}$ and $\Omega_{\beta}$,
\begin{equation}
\langle \Omega_{\alpha}
|Q(A_{t/n})^{n}-Q(e^{-iAt/n})^{n}|\Omega_{\beta}\rangle
e^{-\beta^{*}\beta} \prec n(t/n^{2}).
\end{equation}
By (2.35),
\begin{equation}
\begin{split}
&\langle \Omega_{\alpha} |Q(e^{-iAt/n})^{n}|\Omega_{\beta}\rangle \\
&=\int\prod_{j=1}^{n}\!d\psi_{j}^{*}d\psi_{j}\:
\exp\sum_{j=0}^{n}[(\psi_{j+1}-\psi_{j})^{*}\psi_{j} -
iA(\psi_{j}^{*},\psi_{j})t/n].
\end{split}
\end{equation}
The theorem follows from (3.7), (3.12), and (3.13).
\end{proof}

\begin{remark}
In the notation $\tau_{j}=jt/n,\ \psi_{\tau_{j}} = \psi_{j},\
j=0,1,2,\ldots,n$,
and  $\Delta \tau_{j}= \tau_{j+1}-\tau_{j}$,
$\Delta \psi_{j}= \psi_{j+1}-\psi_{j}$,
the  multiple integral \mbox{(3.2)} is
\[
\int\prod_{j=1}^{n}\!d\psi_{\tau_{j}}^{*}d\psi_{\tau_{j}}\:
\exp i\sum_{j=0}^{n}\Delta \tau_{j}\left[-i\langle\Delta\psi_{\tau_{j}}/
\Delta \tau_{j} | \psi_{\tau_{j}}\rangle -
A(\psi_{\tau_{j}}^{*},\psi_{\tau_{j}})\right].
\]
Its limit at $n=\infty$ (if exists) is the Klauder-Feynman integral
\cite{Kla60} for the coherent states propagator
\[
\langle \Omega_{\alpha} |e^{-iQ(A)t}|
\Omega_{\beta}\rangle e^{-\beta^{*}\beta} =
\int_{\alpha}^{\beta}\prod_{0< \tau < t}\!d\psi_{\tau}^{*}d\psi_{\tau}\:
     \exp i\int_{0}^{t}d\tau
\left[-i\langle \dot{\psi}_{\tau} | \psi_{\tau}\rangle -
A(\psi_{\tau}^{*},\psi_{\tau})\right],
\]
where, similarly to \cite{KD84}, the Hamiltonian functional is
replaced with the antiwick symbol $A$.

   Theorem 3.2  gives sufficient  conditions  for the limit existence,
   even  in the  case of infinite  degrees of freedom.
\end{remark}

As the limit of the multiple functional integrals,
   the Feynman integral over a product Hilbert space is equal to the iterated
   integrals.
Let $c$ be a $|\!|\cdot|\!|_{+}$-continuous
linear canonical transformation
of $\mathcal{H}$ as a real symplectic space. (Unlike \cite{Ber66}
we do not assume that the canonical transformation is proper.)
      Then we have the substitution rule
\[
\begin{split}
&\int_{c\alpha}^{c\beta}\prod_{0< \tau < t}\!d\psi_{\tau}^{*}d\psi_{\tau}\:
     \exp i\int_{0}^{t}d\tau
\left[-i\langle c^{-1}\dot{\psi}_{\tau} | c^{-1}\psi_{\tau}\rangle -
A(c^{-1}\psi_{\tau}^{*},c^{-1}\psi_{\tau})\right]\\
&=\
\int_{\alpha}^{\beta}\prod_{0< \tau < t}\!d\psi_{\tau}^{*}d\psi_{\tau}\:
     \exp i\int_{0}^{t}d\tau
\left[-i\langle \dot{\psi}_{\tau} | \psi_{\tau}\rangle -
A(\psi_{\tau}^{*},\psi_{\tau})\right].
\end{split}
\]
This follows from Theorem 3.1 since the product
$\prod_{j=1}^{n}\!d\psi_{j}^{*}d\psi_{j}$ is a symplectic invariant.

\subsubsection*{Acknowledgment}
I am  grateful to the editors M. Agranovich and M. Shubin for their
inexhaustible patience and to the  referees for their quality control.

\bibliographystyle{amsalpha}

\end{document}